\documentclass[conference,onecolumn,draftcls]{IEEEtran}
\usepackage{graphicx}
\usepackage{amsthm,amsmath}
\usepackage{cite}
\usepackage{url}
\theoremstyle{definition} 

\newcommand{\eq}[1]{\,\begin{equation}
                   #1 
                   \end{equation}
}

\newcommand{\E}[1]{\,E\{#1\}}
\newcommand{\morabba}[1]{\,\begin{flushright}
 \Rectsteel \\
\end{flushright}}

\newcommand{\eqq}[2]{\,\begin{equation} \label{#2}
                   #1 
                   \end{equation}
}

\usepackage{setspace}

\usepackage{caption}
\usepackage{subcaption}

\title{Migration in a Small World: A Network Approach to Modeling Immigration Processes}

\author{\IEEEauthorblockN{Babak Fotouhi and Michael G. Rabbat}
\IEEEauthorblockA{Department of Electrical and Computer Engineering\\
McGill University, Montr\'eal, Qu\'ebec, Canada\\
Email: babak.fotouhi@mail.mcgill.ca, michael.rabbat@mcgill.ca}
}

\begin{document}

\maketitle

\begin{abstract}
Existing theories of migration either focus on micro- or macroscopic behavior of populations; that is, either the average behavior of entire population is modeled directly, or decisions of individuals are modeled directly. In this work, we seek to bridge these two perspectives by modeling individual agents decisions to migrate while accounting for the social network structure that binds individuals into a population.
Pecuniary considerations combined with the decisions of peers are the primary elements of the model, being the main driving forces of migration. 
People of the home country are modeled as nodes on a small-world network. A dichotomous state is associated with each node, indicating whether 
it emigrates to the destination country or it stays in the home country.  We characterize the emigration rate in terms of the relative welfare and population of the home and destination countries. The time 
evolution and the steady-state fraction of emigrants are also derived.
\end{abstract}

\section{Introduction}
There exists no prevalent and coherent theory of international migration yet.
Examples of notable observations in this area include existence of ethnic clustering, 
changes in migration patterns (such as Europe, in the second half of the 20th century) and occasional overflows of immigration volumes (such as Mexico-US) \cite{massey2}. This work considers the structural processes active in emigrant-emanating and immigrant-absorbing countries, and the micro processes motivating 
individuals and enhancing mobility. 

First efforts to lay a theoretical foundation to study emigration date back to 1889 \cite{ravenstein}. Presently, various approaches are segmented by 
difference in perspective, level of analysis and theoretical tools. Neoclassical economics on the macro level reduces migration to wage differentials, 
whose elimination wipes out the migration flow. It makes labor markets the driving mechanisms of international migration, which should be focused on to 
control the in/out-flow of labor \cite{ranis, harris, faini, hatton}. This approach is accompanied by
 equivocal empirical investigations  \cite{lundborg, fields}. 

On the micro level, studies are still mainly based on monetary features, where agents optimize productivity given their skills and necessary costs. The 
expected net return at the outset is modeled as follows \cite{sjaastad, todaro, hatton2, massey1}
\eq{
ER(0)=\int_{t=0}^{n} \bigg[ P_1(t) P_2(t) Y_d(t) - P_3(t) Y_O(t) \bigg] e^{-r t} dt - C(0)
}
where $P_1(t)$ is the probability of evading deportation from the destination country, $P_2(t)$ is the employment probability at the destination, $Y_d(t)$ is the 
average wage there, $P_3(t)$ is the employment probability in the home country,  $Y_O(t)$ is the average earning in the home country, $r$ is the rate of inflation, $n$ is the time horizon, and $C(0)$ is the value of total migration costs
 at the outset. Migration goes on as long as the expected net return is positive. This model is supported by empirical evidence \cite{bowles, straubhaar, massey2}. 
 
Both of the approaches described above fail to capture the topology of the social network people live in. Another potent factor is the stock of immigrants already present at the destination, providing beneficial network externalities (the influence of social peers), performing as conduits of information 
and support, facilitating adjustment to the new place \cite{epstein, massey2, darvish, bauer1, haung}. 
This feature is backed by ample empirical evidence \cite{munshi, bauer2, pedersen, dunlevy, jaeger, mckenzie}. Network
externality  and cascading (or herding) effect are two inevitable elements of the network perspective. 

In this contribution we approach migration processes from the perspective of network science. We study the effect of social influence and wage differentials on the fraction of the population who emigrate. We also introduce a bias towards the home land, to model the inherent patriotism people usually have, and find its effect on emigration rates.

\section{The Model} 

To model social interactions within the population of a country, we employ a simple small-world network model~\cite{watts}. At first, nodes are configured symmetrically  on a circle, and each node is connected to every node within the proximity of $b$ 
steps away from it, giving them an initial degree of $2b$. Then each pair with distance more than $b$ is connected with probability $p$. 
The expected total number of links is then $ Nb+ \frac{N(N-1-2b)}{2}p$  and the average degree of nodes will be $2b+\Theta(Np)$. To ensure that the average degree is finite 
in the limit as $N \rightarrow \infty$ it is required that  $p\sim \frac{1}{N}$.

Each node $x$ is endowed with a state $s_x(t) \in \{+1, -1\}$, indicating whether the node stays in its home country ($+1$) or emigrates to the destination country ($-1$). The states evolve over time according to dynamics which depend on factors discussed next. 
At each time step, a node observes the state of the adjacent nodes and finds the fraction of those that are $+1$ or $-1$. 
Denote the fraction of the entire population who, at $t$, will stay home (with $s_x(t)=+1$) by $\rho(t)$; emigrants (with $s_x(t) =-1$) are the other $1-\rho(t)$ portion. Each node also observes the expected wage in the home country and the foreign country. 
Here we consider a simplified case where the wage of a country equals a characteristic indicator of its national wealth, denoted by $G$, divided by its population. 
We denote the population of the home and destination country by 
$N_h$ and $N_d$ respectively, and define $\alpha \stackrel{\text{def}}{=} \frac{N_d}{N_h}$.

\subsection{Migration Without Patriotic Bias}

We first develop a model in which agents make emotionless, rational decisions solely on social interactions and pecuniary considerations (sometimes referred to as \emph{homo economicus}), without any bias towards their homeland. 

The expected wage per capita for the home country is 
\eq{
w_+(t)= \frac{G_h}{N_h \rho(t)}
}
and for the destination country it equals
\eq{
w_-(t)=\frac{G_d}{N_d+N_h(1-\rho(t))}
}
 If one was to only consider these two criteria (wages in the home and destination countries), the probabilities corresponding to leaving and staying would become
\eq{
\begin{cases}
\displaystyle r_+(t)=\frac{w_+(t)}{w_+(t)+w_-(t)}=  \frac{\frac{G_h}{N_h \rho(t)}}{\frac{G_h}{N_h \rho(t)}+\frac{G_d}{N_d+N_h(1-\rho(t))}} = \frac{G_h (\alpha+1-\rho(t))}{G_h(\alpha+1-\rho(t))+G_d \rho(t)}\\ \\
\displaystyle r_-(t)=\frac{w_-(t)}{w_+(t)+w_-(t)}= \frac{\frac{G_d}{N_d+N_h(1-\rho(t))}}{\frac{G_h}{N_h \rho(t)}+\frac{G_d}{N_d+N_h(1-\rho(t))}} =\frac{G_d \rho(t)}{G_h(\alpha+1-\rho(t))+G_d \rho(t)}
\end{cases}
}

We also wish to model social influence. Let $\mathcal{N}_x$ denote the set of neighbors of node $x$, and let $z_x = |\mathcal{N}_x|$ denote the degree of node $x$. We model social influence in a manner akin to the conventional voter model~\cite{liggett}. Each node adjusts its decision (in expectation) according to the fraction of neighbors who are staying (i.e., those neighbors $y \in \mathcal{N}_x$ with $s_y(t) = +1$) and those who are leaving (i.e., $s_y(t) = -1$). The probabilities it assigns to the actions of staying or going are proportional to these fractions. Combining this with the wage considerations, for node $x$ we get the following decision probabilities
\eqq{
\begin{cases}
\displaystyle P\{s_x(t+\Delta t) = +1 \} = \theta \left[ \frac{\sum_{y\in \mathcal{N}_x} \delta(s_y(t),1)}{z_x}\right] + (1-\theta) r_+(t) \\ \\
\displaystyle P\{s_x(t+\Delta t) = -1 \} = \theta \left[ 1-\frac{\sum_{y\in \mathcal{N}_x} \delta(s_y(t),1)}{z_x}\right] + (1-\theta) r_-(t)
\end{cases}
}{pp1}
where $\theta \in (0,1)$ is a parameter quantifying the weight given to network effects vs.~monetary considerations, and $\delta(a,b)$ is equal to unity when the arguments match and is zero otherwise. 

We would like to characterize the dynamics of $\rho(t)$. To do this, we study how the average state evolves with time. The expected value of the state for node $x$ in the next timestep is
\eq{
\E{s_x(t+\Delta t)} = (+1)  P\{s_x(t+\Delta t) = +1 \} + (-1)  P\{s_x(t+\Delta t) = -1 \} = 2 P\{s_x(t+\Delta t) = +1 \} -1.
}
Using (\ref{pp1}) we get 
\eqq{
\E{s_x(t+\Delta t)} = 
2\theta \left[ \frac{\sum_{y\in N_x} \delta(s_y,1)}{z_x}\right] + 2(1-\theta) \frac{G_h (\alpha+1-\rho(t))}{G_h(\alpha+1-\rho(t))+G_d \rho(t)} -1.
}{Es1}
Now we need to derive an expression for the first term. Let $A$ denote the adjacency matrix of the network. Then we have
\eq{
\frac{\sum_{y\in N_x} \delta(s_y,1)}{z_x} = \sum_{y \neq x} \frac{A_{xy} \delta(s_y,1)}{z_x}.
}
In the so-called annealed approximation \cite{vilone}, one studies quantities averaged over the ensemble of all 
graph realizations (where the average of some quantity $\psi$ denoted by $\langle \psi \rangle$)
, and approximates $ \frac{A_{xy} \delta(s_y,1)}{z_x}$ by $\langle \frac{A_{xy}}{z_x} \rangle \langle \delta(s_y,1) \rangle$.
%
All $A_{ij}$s are random variables with binary distribution, and the average is over all configurations. We need to consider two different cases, depending on whether or not $x$ and $y$ are initially within $b$ hops of each other in the small-world model, before long-range connections are added. If $y$ is one of the $2b$ initial neighbors of $x$, then 
\eq{
\left\langle \frac{A_{xy}}{z_x} \right\rangle = \sum_{k=0}^{N'} \frac{P\{ z_x=k+2b \} }{k+2b}=\sum_{k=0}^{N'} \frac{\binom{N'}{k}p^k(1-p)^{N'-k} }{k+2b} \stackrel{\text{def}}{=} \omega,
}
where $N'=N-2b-1$ is the maximum number of links which may be added, per node, in the small-world model.
If $y$ is not initially linked to $x$ then we have
\eq{
\left\langle \frac{A_{xy}}{z_x} \right\rangle = p\sum_{k=0}^{N'-1} \frac{P\{ z_x=k+2b+1 \} }{k+2b+1}=p \sum_{k=0}^{N'-1} \frac{\binom{N'-1}{k}p^k(1-p)^{N'-k-1} }{k+2b+1} \equiv \Omega
.}

As noted in \cite{vilone}, one can express $\omega$ as the following integral
\begin{align}
\omega &=  \sum_{k=0}^{N'} \frac{P\{ z_x=k+2b \} }{k+2b} \\
&=\sum_{k=0}^{N'} \binom{N'}{k}(1-p)^{N'-k}\frac{p^k }{k+2b} \\
&= \frac{1}{p^{2b}}\sum_{k=0}^{N'} \binom{N'}{k}(1-p)^{N'-k}\frac{p^{k+2b} }{k+2b} \\
&= \frac{1}{p^{2b}}\sum_{k=0}^{N'} \binom{N'}{k}(1-p)^{N'-k} \int_{\lambda=0}^{p} \mathbf{d}{\lambda}~ \lambda^{k+2b-1}  \\
&= \frac{1}{p^{2b}} \int_{\lambda=0}^{p} \mathbf{d}{\lambda} ~\lambda^{2b-1} \left[ \sum_{k=0}^{N'} \binom{N'}{k}(1-p)^{N'-k} \lambda^k\right] \\
&= \frac{1}{p^{2b}} \int_{\lambda=0}^{p} \mathbf{d}{\lambda} ~\lambda^{2b-1} \left[ (\lambda+1-p)^{N'}\right],
\end{align}
and similarly
\eq{
\Omega= \frac{1}{p^{2b}} \int_{\lambda=0}^{p} \mathbf{d}{\lambda} ~\lambda^{2b} \left[ (\lambda+1-p)^{N'-1}\right]
.}

Integrating by part gives
\begin{align}
\omega &= 
 \frac{1}{p^{2b}} \int_{\lambda=0}^{p} \mathbf{d}{\lambda} ~\lambda^{2b-1} \left[ (\lambda+1-p)^{N'}\right] \\
 &= 
\frac{1}{p^{2b}} \left[ \frac{p^{2b}}{2b}-\frac{N'}{2b} \int_{\lambda=0}^{p} \mathbf{d}{\lambda} ~\lambda^{2b} (\lambda+1-p)^{N'-1} \right] \\
&=\frac{1}{2b}- \frac{N-2b-1}{(2b)p^{2b}} \int_{\lambda=0}^{p} \mathbf{d}{\lambda} ~\lambda^{2b} (\lambda+1-p)^{N'-1}\\
 &=\frac{1}{2b}- \left( \frac{N-2b-1}{2b} \right) \Omega,
\end{align}
from which it is clear that
\eq{
2b\omega + \Omega (N-2b-1) =1.
}
Combining this with (\ref{Es1}) we get
\eq{
2 \rho(t+\Delta t)-1 = 2 \theta \rho(t) - 1 + 2(1-\theta) \frac{G_h (\alpha+1-\rho(t))}{G_h(\alpha+1-\rho(t))+G_d \rho(t)}
}
which leads to the following differential equation for $\rho$
\eq{
\dot{\rho} = (\theta -1) \rho + (1-\theta) \frac{G_h (\alpha+1-\rho)}{G_h(\alpha+1-\rho)+G_d \rho}.
}

Note that this differential equation is separable, and so its solution is expressed as
\eq{
(1-\theta) t = \int_{\rho(0)}^{\rho(t)} \mathbf{d}\rho \left[ - \rho + \frac{G_h (\alpha+1-\rho)}{G_h(\alpha+1-\rho)+G_d \rho} \right]^{-1}
}
The closed form expression for this integral is given by
\begin{align}
 - \frac{1}{2} & \ln \left[ \frac{(1+\alpha)G_h(1-\rho(t)) - \rho(t)(1-\rho(t))G_h-G_d \rho(t)^2}{-G_d} \right] \nonumber \\
&+\frac{\alpha \sqrt{G_h} \tanh^{-1} \left[ \frac{G_h (\alpha + 2-2\rho(t)) + 2G_d \rho(t)}{\sqrt{G_h} \sqrt{G_h (1+\alpha)^2+4(1+\alpha)G_d-G_h(1+2\alpha)}} \right]
- \alpha \sqrt{G_h} \tanh^{-1} \left[ \frac{G_h \alpha + 2 G_d }{\sqrt{G_h} \sqrt{G_h (1+\alpha)^2+4(1+\alpha)G_d-G_h(1+2\alpha)}} \right]  }
{\sqrt{G_h (1+\alpha)^2+4(1+\alpha)G_d-G_h(1+2\alpha)}}  
\end{align}
which is not very informative. Thus we seek the steady-state solution which can be found by setting the time derivative equal to zero, which gives
\eq{
\rho_\infty \stackrel{\text{def}}{=} \lim_{t\rightarrow \infty} \rho(t) =\frac{\sqrt{(\alpha+2)^2+4G_h(G_d-G_h)(1+\alpha)}-(\alpha+2)}{2(G_d-G_h)}.
}

\begin{figure}[ht]
  \centering
  \includegraphics[width=4in]{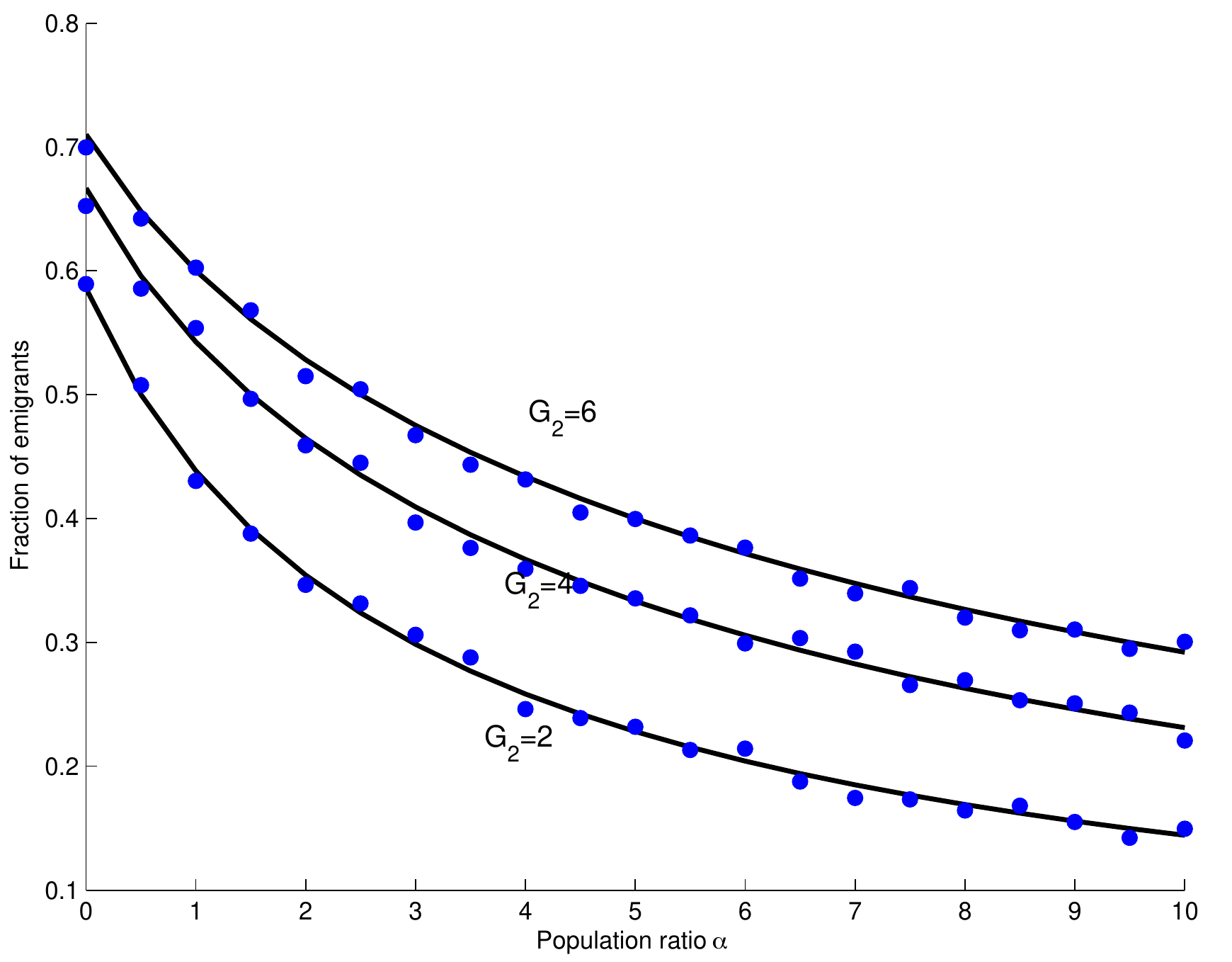}
  \caption[Figure ]%
  {The steady-state fraction of the population who decide to leave ($1 - \rho_\infty$), as a function of $\alpha$, for different values of $G_d$, when $G_h$ is normalized to unity, and $\theta=0.7$. There are 
$N=1000$ nodes, with $b=8$  and $p=4/N$, the steady-state is acquired after 2000 timesteps. 
Solid lines and bullets represent theoretical prediction and simulation results respectively. }
\label{fig1}
\end{figure}

\begin{figure}[ht]
  \centering
  \includegraphics[width=4in]{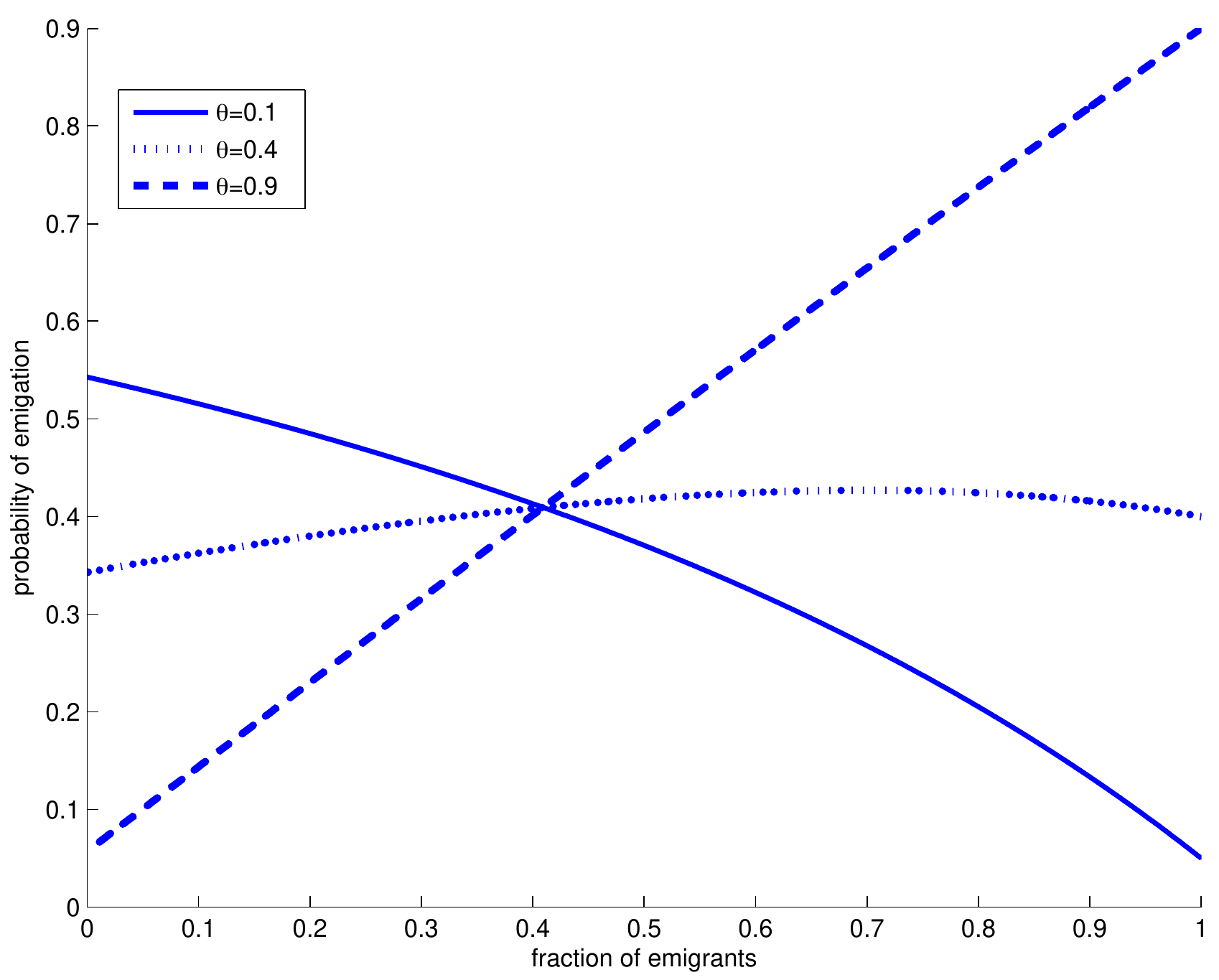}
  \caption[Figure ]%
  {Emigration probability $P\{s_x(t+\Delta t) = - 1\}$ as a function of the fraction of those already emigrated, $1-\rho(t)$, for different values of $G_d$, when $G_h = 1$. There are 
$N=1000$ nodes, with $b=8$  and $p=4/N$.}
\label{fig2}
\end{figure}

We validate the expressions derived above via simulation. Figure \ref{fig1} depicts the equilibrium fraction of emigrants as a function of $\alpha$ for different values of $G_d$ when $G_h = 1$. Figure \ref{fig2} shows the expected probability of emigration as a function of the fraction of those who already migrated.

\subsection{Migration With Patriotic Bias}

Note that so far there is no intrinsic tendency towards ``home'', homo-economicus-type agents are in action and high probabilities are not surprising. For instance, this helps explain why the fraction of nodes emigrating can reach 70\% in Figure~\ref{fig1}.

To be more realistic, one must account for the disposition against emigrating due to patriotism. We account for this in a Bayesian manner. Let $1 - \gamma$ denote the prior probability of staying at home regardless of social and economic conditions. Incorporating this factor into the model described in the previous section leads to the expected state dynamics
\begin{align}
\E{s_x(t+\Delta t)} &= \gamma \bigg[ (+1)  P\{s_x(t+\Delta t) = + 1\} + (-1)  P\{s_x(t+\Delta t) = - 1\} \bigg] + (1-\gamma)(+1) \\
&= \gamma \bigg[ 2 P\{s_x(t+\Delta t) = + 1\} -1 \bigg] + 1-\gamma.
\end{align}
The equation of motion becomes
\eq{
\dot{\rho} = (\gamma \theta -1) \rho + \gamma (1-\theta) \frac{G_h (\alpha+1-\rho)}{G_h(\alpha+1-\rho)+G_d \rho} + 1-\gamma.
}
The solution to this differential equation can be found by taking the integral
\eq{
t= \int \frac{d \rho(t)}{(1-\gamma)+\rho(t)(\gamma \theta -1) + \gamma (1-\theta) \frac{G_h (\alpha+1-\rho(t))}{G_h(\alpha+1-\rho(t))+G_d \rho(t)}}
}
which is straightforward but voluminous. One can simplify the above expression when the fraction of emigrants is very small\footnote{According to the CIA World Fact Book, the vast majority of countries have migration rates below 1\%. \\\url{https://www.cia.gov/library/publications/the-world-factbook/rankorder/2112rank.html}} by setting $\rho(t)= 1-\epsilon$, and taking a Taylor series expansion about $\epsilon=0$ up to second order terms. Defining
\eq{
\begin{cases}
 a \equiv \gamma (1-\theta) \left( \frac{G_d}{\alpha G_h + G_d}\right) \\
b \equiv 1-\gamma \theta + \frac{\gamma(1-\theta)G_h}{G_h \alpha+G_d} \left( 1+\frac{\alpha(G_d-G_h)}{G_h \alpha + G_d} \right) \\
c \equiv \frac{\gamma (1-\theta) G_h (G_h-G_d) (\alpha-1)}{(G_h \alpha +G_d)^2}
\end{cases}
}
we have
\eq{
t=\frac{1}{a} \left[ \epsilon - \frac{b}{2a} \epsilon^2 + \left( \frac{b^2}{3a^2}-\frac{c}{3a}\right) \epsilon^3  + O(\epsilon^4)\right].
}

Alternatively, the equilibrium emigration fraction is found by solving 
\eq{
(\gamma \theta -1) \rho(t) + \gamma (1-\theta) \frac{G_h (\alpha+1-\rho(t))}{G_h(\alpha+1-\rho(t))+G_d \rho(t)} + 1-\gamma = 0
}
whose solution is 
\eq{
\rho_\infty \stackrel{\text{def}}{=} \lim_{t \rightarrow \infty} \rho(t) =\frac{-B - \sqrt{B^2-4AC}}{2A}
}
with
\eq{
\begin{cases}
 A \equiv (1-\gamma \theta)(G_h-G_d) \\
B \equiv (\alpha+1)G_h (\gamma \theta-1) + (1-\gamma)(G_d-G_h)-\gamma G_h (1-\theta) \\
C \equiv G_h (1+\alpha)(1-\gamma \theta).
\end{cases}
}
Note that the sign of $\rho_\infty$ depends on the sign of $A$; if $G_d > G_h$, then $\rho_\infty > 0$ and there is emigration to the destination country, as one would expect.

\begin{figure}[ht]
  \centering
  \includegraphics[width=4in]{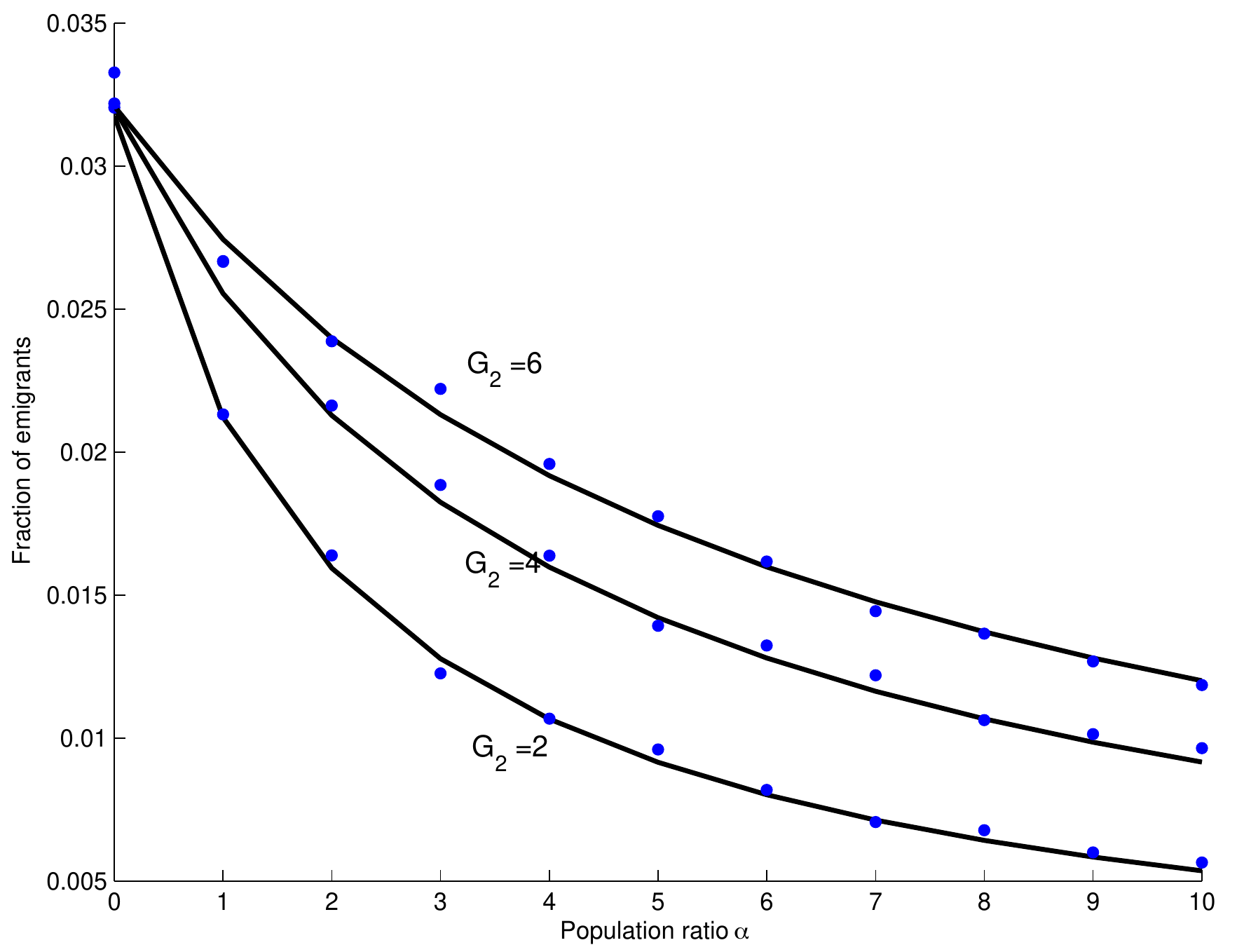}
  \caption[Figure ]%
  {The fraction of the population who decide to leave, as a function of $\alpha$, for different values of $G_d$, when $G_h$ is normalized to unity,
 with $\theta=0.7$,  $\gamma=0.2$, $N=1000$, $b=8$ and $p=4/N$. Solid lines and bullets represent theoretical prediction and simulation results respectively.}
\label{fig3}
\end{figure}

\begin{figure}[ht]
  \centering
  \includegraphics[width=4in]{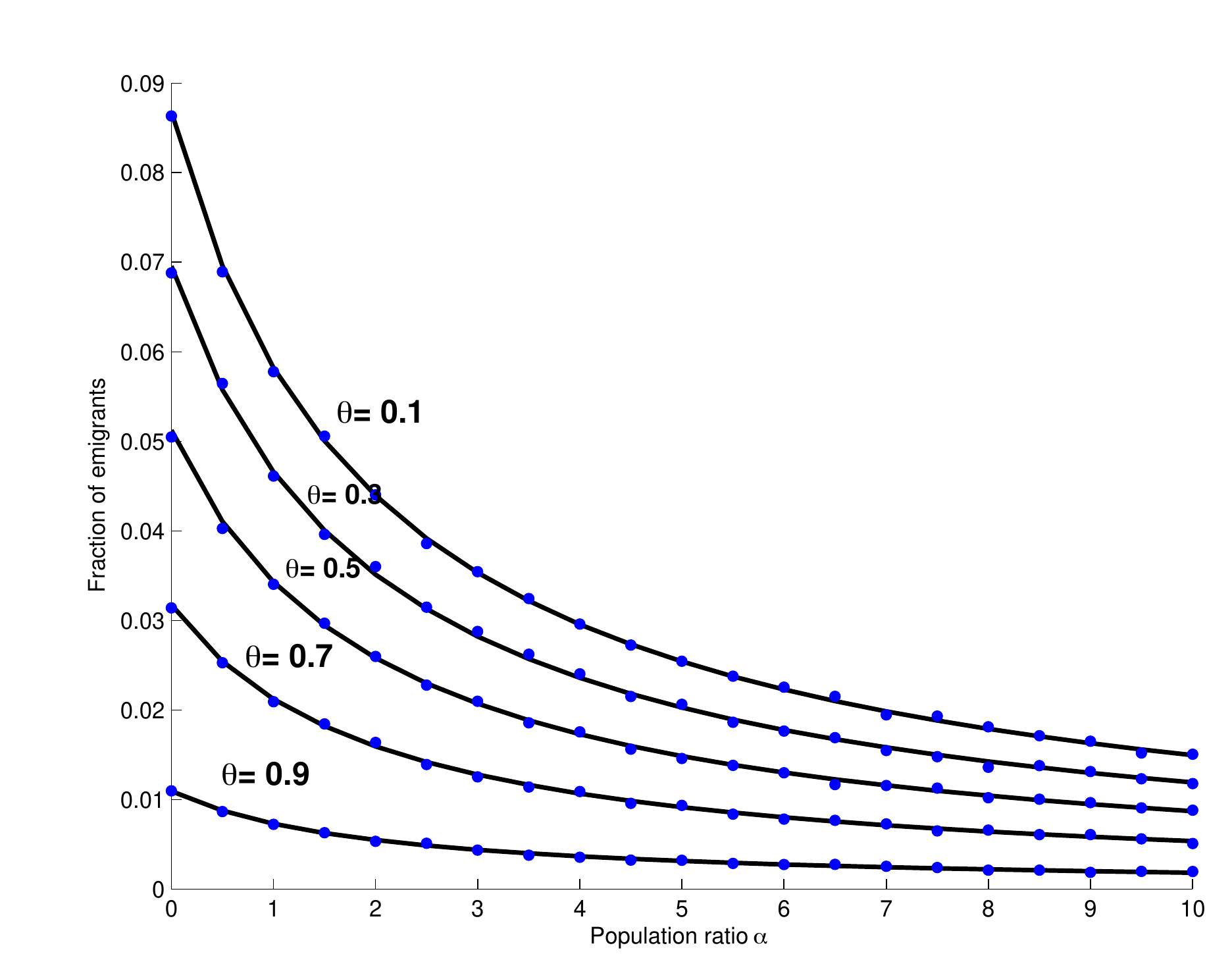}
  \caption[Figure ]%
  {The fraction of the population who decide to leave, as a function of $\alpha$, for different values of $\theta$, 
when $G_h$ is normalized to unity, with $G_d=2$, $\gamma=0.2$, $N=1000$, $b=8$  and $p=4/N$. 
Solid lines and bullets represent theoretical prediction and simulation results respectively.}
\label{fig4}
\end{figure}

\begin{figure}[ht]
  \centering
  \includegraphics[width=4in]{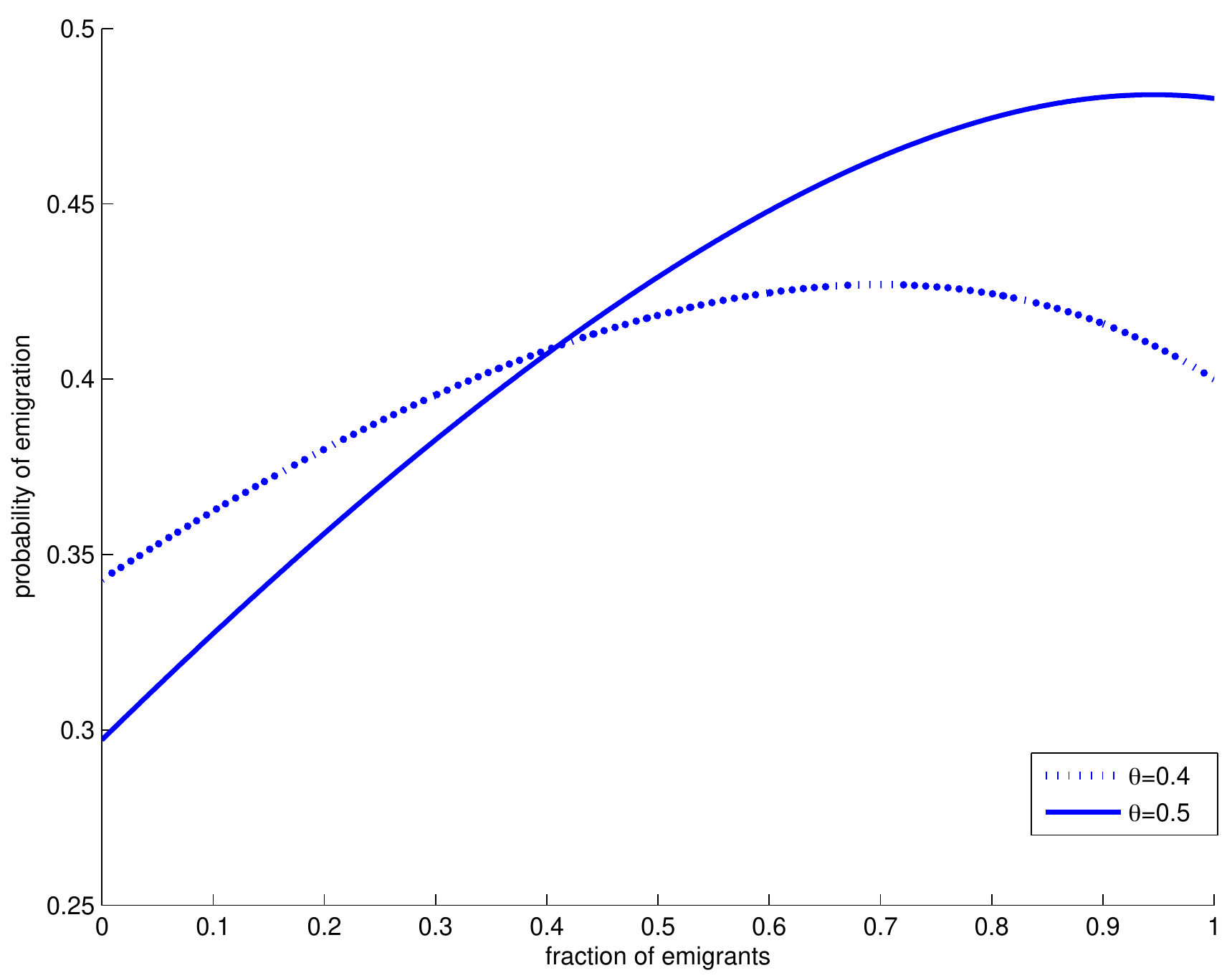}
  \caption[Figure ]%
  {Emigration probability , as a function of $1-\rho(t)$, the fraction of those already emigrated, for different values of $G_d$, when $G_h$ is normalized to unity, with $\gamma=0.2$. }
\label{fig5}
\end{figure}

The equilibrium fraction of emigrants is depicted in Figures \ref{fig3} and \ref{fig4}, and the emigration probability with respect to 
the fraction of those already done so is given in figure (\ref{fig5}), which replicates the one due to network externalities and herding effect in \cite{bauer1}. 

Note that in the calculations, the role of $p$ (the connection probability  the small-world model)  cancelled out. In the simulations no effect on the emigration fractions was observed by changing $p$. This peculiarity of the small-world network coincides with the findings in \cite{vilone}, where the average magnetization is preserved and the correlation length of the finite graph reaches unity in the steady state, irrespective of the value of $p$. When there is perfect symmetry between all nodes, local densities vanish in the long time limit, and the equilibrium fraction of emigrants (or in the voter model as in \cite{vilone}, average magnetization) only depends on the total fraction of the emigrants. 
This suggests that, in order to further incorporate the structural properties of the society, one must introduce asymmetry between nodes by construction, that is, enforce hierarchy of influence or stratify the underlying network.

\begin{figure}[ht]
  \centering
  \includegraphics[width=4in]{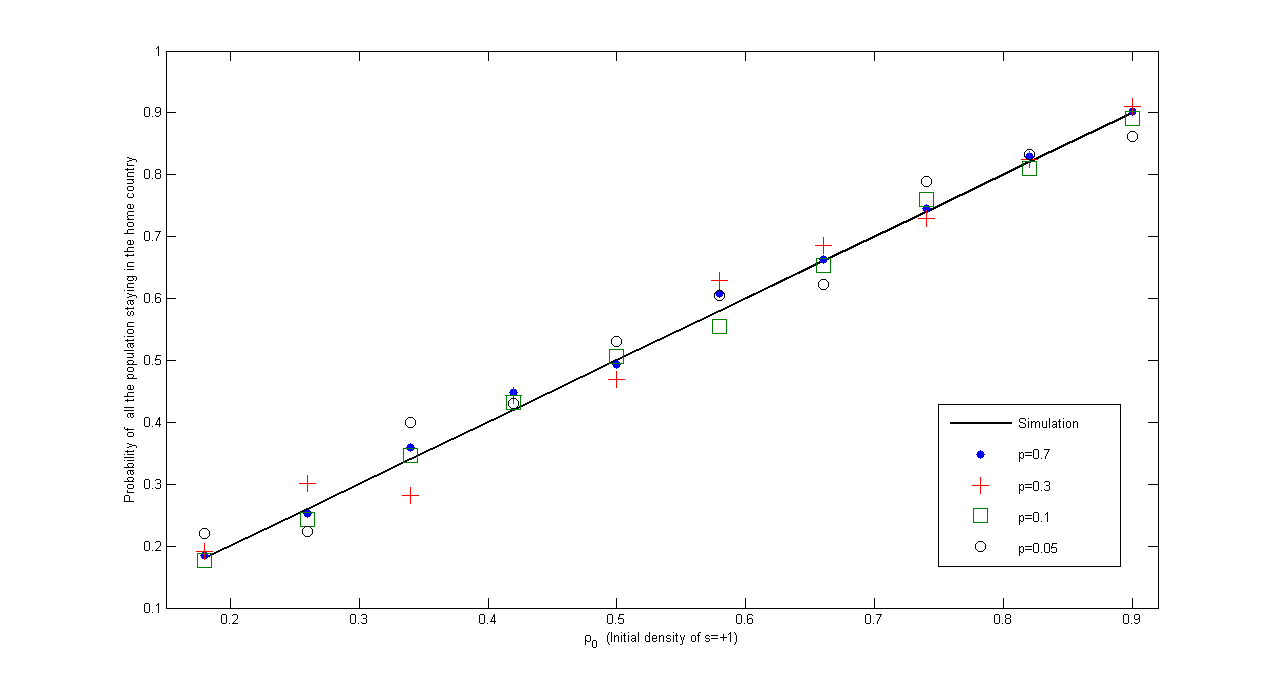}
  \caption[Figure ]%
  {The fraction of the population who stay in the home country, as a function of those decided to do so at the outset, for different values of $p$, the connection probability of the small-world graph.  }
\label{fig_voter}
\end{figure}

Also note that, in the case of $\gamma=\theta=1$ we have the conventional voter model, which only has two absorbing states; either all nodes end up with $s=+1$ or with $s=-1$. Figure~\ref{fig_voter} represents the probability of al the population staying in the home country, as a function of those who decided to do so at the outset. Note that the results are the same for different values of $p$.

\begin{figure}[ht]
       
        \begin{subfigure}[b]{0.7\textwidth}
                \centering
                \includegraphics[width=\textwidth]{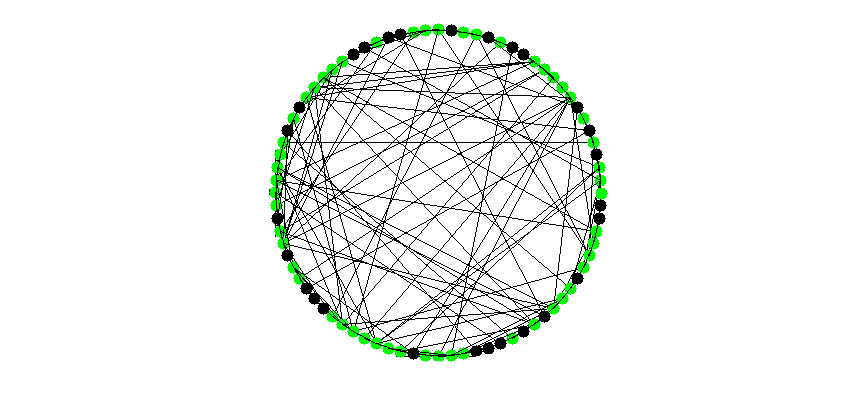}
                \caption{$\theta=0.2$ }
                \label{fig_TH1}
        \end{subfigure}%
        \!\!\!\!\!\!\!\!\!\!\!\!\!\!\!\!\!\!\!\!\!\!\!\!\!\!\!\!\!\!\!\!\!\!\!\!
        \!\!\!\!\!\!\!\!\!
       \begin{subfigure}[b]{0.5\textwidth}
                \centering
                \includegraphics[width=\textwidth]{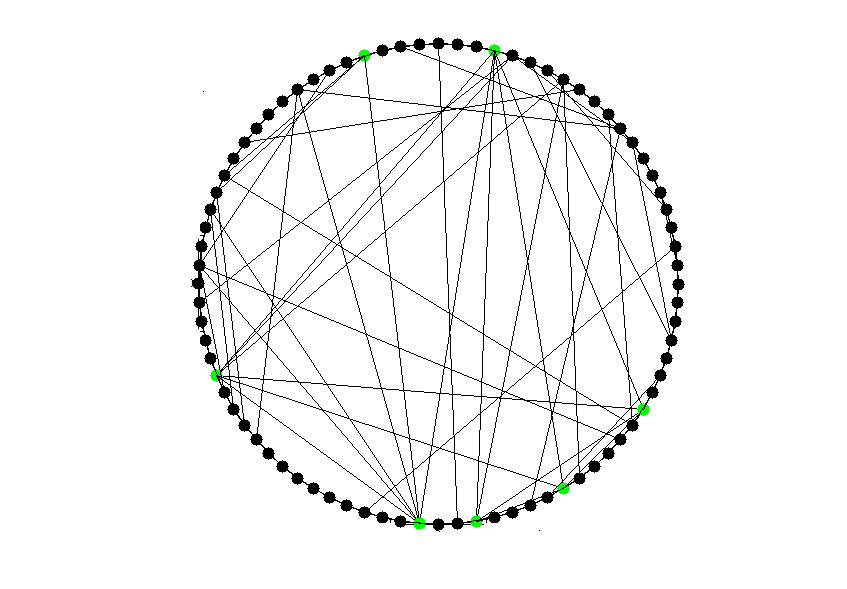}
                \caption{$\theta=0.8$ }
                \label{fig_TH2}
        \end{subfigure}%
     \caption{The small-world graph in the steady-state. Green nodes are the emigrants and black nodes are those who stay in the home country. $N=80$, $p=\frac{2}{N}$, $\gamma=0.7$, $G_2=2$ and $\alpha=1.5$. For the graph on the left, the value of $\theta$ is set to $0.2$, and for the one on the right, it is set to $0.8$.} \label{fig}
\end{figure}

Figures ~\ref{fig_TH1} and ~\ref{fig_TH2} show the effect of $\theta$ on the fraction of emigrants. The network has $N=80$ nodes with connection probability $p=2/N$, and all other parameters rather than $\theta$ are the same for the two graphs.

%
%

\section{Summary}

Previous models proposed for the emigration process do not take into account the social network structure of the population. In this paper we have modelled emigration decisions taken by agents interacting over a small-world network. The first model was merely utility based. The fraction of emigrants were found in terms of the difference in average wage in the home and destination countries, the population ratio of the population of the two countries, and the relative importance that nodes give to their personal utility calculations and to imitating the behavior of others. 
The second model embodied a default disposition against leaving the homeland, modelling patriotism. Theoretical predictions were accompanied by simulations. We also found that as long as the nodes are indistinguishable, i.e. their position and characteristics in the network are identical, then the average emigration rates are independent of the connection probability in the small-world network. One must introduce hierarchy or other structural asymmetries in order to study the dependence of the collective behavior on the structural properties of the underlying network.


\begin{thebibliography}{99}



\bibitem{ravenstein}
E. Ravenstein, ``The Laws of Migration'', Journal of Royal Statistical Society, Vol. 52 (1889), pp. 241-305.
\bibitem{ranis}
G. Ranis, J. C. H. Fei, ``A Theory of Economic Development'', American Economic Review, Vol. 53 (1961), pp. 533-565.
\bibitem{harris}
J. R. Harris, M. P. Todaro, ``Migration, Unemployment, and Development: A Two-sector Analysis'', American Economic Review, Vol. 60 (1970), pp. 126-142.
\bibitem{faini}
R. Faini, A. Venturini, ``Migration and Growth: The Experience of Southern Europe'', CEPR, London (1994). 
\bibitem{hatton}
T. Hatton, J. G. Williamson, ``Latecomers to Mass Emigration'', The Migration and the International 
Labor Market 1850-1939, Routledge, London (1994)
\bibitem{lundborg}
P. Lundborg, ``Determinants of Migration in the Nordic Labor Market'', Scandinavian Journal of Economics, Vol. 93 (1991), pp. 363-375. 
\bibitem{fields}
G. S. Fields, ``Place-to-place Migration: Some New Evidence'', The Review of Economics and Statistics, Vol. 61 (1991), pp. 21-32.



\bibitem{sjaastad}
L. Sjaastad, ``The Costs and Returns of Human Migration'', Journal of Political Economy, Vol. 70, N. 5 (1962), pp. S80-S93.
\bibitem{todaro}
M. P. Todaro, ``A Model of Labor Migration and Urban Unemployment in Less-developed Countries'', The American Economic Review, Vol. 59 (1969), pp. 138-148. 
\bibitem{hatton2}
T. Hatton, J. G. Williamson, ``What Fundamentals Drive World Migration?'', NBER Working Paper No. 9159 (2002). 
\bibitem{massey1}
D. S. Massey et al. , ``Theories of International Migration: A Review and Appraisal'', Population and Development Review, Vol. 19, N. 3 (1993), pp. 431-466.
\bibitem{bowles}
S. Bowles, ``Migration as Investment: Empirical Tests of the Human Investment Approach to Geographical Mobility'', The Review of Economics and Statistics, Vol. 52 (1970), pp. 356-362.
\bibitem{straubhaar}
T. Straubhaar, ``On the Economics of International Migration'', Verlag Paul Haupt, Bern (1988). 
\bibitem{massey2}
D. S. Massey, F. G. Espana, ``The Social Process of International Migration'', Science, Vol. 237, N. 4816 (1987), pp. 733-738.


\bibitem{epstein}
G. S. Epstein, A. L. Hilmann, ``Herd Effects and Migration'', DEPR Discussion Paper No. 1811, London.
\bibitem{darvish}
T. Darvish-Lecker, ``Externalities in Migration'', Economic Letters Vol. 33, Is. 2 (1990), pp. 185-191. 
\bibitem{bauer1}
T. Bauer, G. Epstein, I. B. Gang, ``Herd Effects or Migration Networks? The Location Choice of Mexican Immigrants in the US'', IZA Discussion Paper No. 551 (2002)
\bibitem{epstein}
G. S. Epstein, ``Informal Cascades and Decision to Migrate'', IZA Working Paper, No. 445 (2002). 
\bibitem{haung}
S. Haung, ``Migration Networks and Migration Decision-Making'', Journal of Ethnic and Migration Studies, Vol. 34, No. 4 (2008), pp. 585-605. 
\bibitem{munshi}
K. Munshi, ``Identification of Network Effects: Mexican Migrants in the US Labor Market'', NEUDC Conference, Boston (2001). 
\bibitem{bauer2}
T. Bauer, G. Epstein, I. B. Gang, ``What are Migration Networks?'', IZA Discussion Paper No. 200 (2000). 
\bibitem{pedersen}
P. J. Pedersen, M. Pytlikova, N. Smith, ``Selection or Network Effects? Migration Flows into 27 OECD Countries, 1990-2000'', IZA Discussion Paper, N0. 1104. 
\bibitem{dunlevy}
J. A. Dunlevy, ``On the Settlement Patterns of Recent Carribean and LAtin Immigrants to the United States'', Growth and Change, Vol. 22, Is. 1 (1991), pp. 54-67. 
\bibitem{jaeger}
D. A. Jaeger, ``Local Labor Markets, Admission Categories, and Immigrant Location Choice'', Hunter College and Graduate School CUNY Working Paper (2000). 
\bibitem{mckenzie}
D. McKenzie, ``Self-Selection Patterns in Mexico-U.S Migration: The Role of Migration Networks'', The Review of Economics and Statistics, Vol. 92, No. 4 (2010), pp. 811-821. 



\bibitem{watts} 
D. J. Watts, H. Strogatz, ``Collective Dynamics of 'Small-world' Networks'', Nature Vol. 393, pp. 440-442 (1998)
%
%
%

\bibitem{vilone}
D. Vilone, C. Castellano, ``Solution of Voter Model Dynamics on Annealed Small-world Networks, Physical Review E, V. 69, Is. 1 (2004)

\bibitem{liggett}
T. M. Liggett, ``Interacting Particle Systems'', Springer Verlag, New York, 2005. 


\end{thebibliography}
\end{document}